\documentclass[lengthcheck,final,twocolumn,letterpaper,balancelastpage,prl,showpacs,showkeys,superbib]
{revtex4}
\usepackage{amsmath} 
\usepackage{amsfonts}  
\usepackage{amssymb}  
\usepackage{longtable}  
\usepackage{graphicx}
\usepackage[squaren]{SIunits}

\def\dd{{\rm d}}

\def\etc{{\em{etc.}}}
\def\eg{{\em{e.g.}}}

\def\figsize{0.8}
\newcommand{\ve}{\varepsilon}
\newcommand{\vare}{\varepsilon_{\sf e}}
\newcommand{\varm}{\varepsilon_{\sf m}}
\newcommand{\vari}{\varepsilon_{\sf i}}
\def\dotted{\protect\mbox{-\,-\,-\,-\,-}}
\def\kesik{\protect\mbox{--\, --\, --}}
\def\full{\protect\mbox{--------}}

\begin{document}

\title{The Landau-Lifshitz/Looyenga dielectric mixture expression and its self-similar fractal nature}
\author{Enis Tuncer}
\affiliation{{\tt enis.tuncer@physics.org}}

\begin{abstract}
In this paper, dielectric permittivity of dielectric mixtures is discussed in view of the spectral density representation method. A distinct representation is derived for predicting the dielectric properties, permittivities $\varepsilon$, of mixtures. The peculiar presentation is based on the scaled permittivity $\xi=(\vare-\varm)(\vari-\varm)^{-1}$, where the subscripts `{\sf e}', `{\sf m}' and `{\sf i}' denote the dielectric permittivities of the effective, matrix and inclusion media, respectively [Tuncer E 2005 {\em J. Phys. Condens. Matter} {\bf 17} L125]. This novel form of representation is the same as the distribution of relaxation times formalism in dielectric relaxation. Consequently, we propose an expression for the scaled permittivity, which is the same as one of the extensively used dielectric dispersion expressions, known as the Havriliak-Negami empirical formula. The scaled permittivity representation has potential to be improved and to be implemented in to the existing analyzing routines for dielectric relaxation data to extract the topological/morphological description in mixtures. In order to illustrate the strength of the representation and confirm the proposed hypothesis, Landau-Lifshitz/Looyenga expression is selected, and the structural information of the mixture is extracted. Both a recently developed numerical method to solve inverse integral transforms and the proposed empirical scaled permittivity expression are employed to estimate the spectral density function of the Landau-Lifshitz/Looyenga expression. In the simulations the concentration $q$ of the inclusions phase  are varied. The estimated spectral functions for the mixtures with different inclusion concentration compositions show similar spectral density functions, composed of couple of bell-shaped distributions, with coinciding peak locations. We think therefore that the coincidence is an absolute illustration of a self-similar fractal nature of mixture topology (structure) for the considered Landau-Lifshitz/Looyenga expression. Consequently,  the spectra are not altered significantly with increased filler concentration level--exhibit a self-similar spectral density functions for different concentration levels. Last but not least, the calculated percolation strengths also confirm the fractal nature of the systems characterized by the Landau-Lifshitz/Looyenga mixture expression. We conclude that  the Landau-Lifshitz/Looyenga expression is therefore suitable for complex composite systems that have hierarchical order in their structure, which confirms the finding in the literature.
\end{abstract}
\keywords{
   {Dielectric properties of solids and liquids} ,
   {Optical properties of bulk materials and thin films} ,
   {Permittivity (dielectric function)} ,
   {Composite materials} ,
   {Spectral methods} ,
   {Applications of Monte Carlo methods} ,
   {Fractals} ,
   {Data analysis: algorithms and implementation; data management} ,
   {Disordered solids}
 }
\pacs{
   {77.22.-d}, 
   {78.20.-e}, 
   {77.22.Ch}, 
   {77.84.Lf}, 
   {02.70.Hm}, 
   {02.70.Uu}, 
   {05.45.Df}, 
   {07.05.Kf}, 
   {61.43.-j}
 }

\maketitle


\section{Introduction}
Electrical properties of composite materials have attracted researchers to seek a relation between overall composite properties and intrinsic properties of the parts forming the mixture  (constituents) and their spatial arrangement inside the mixture~\cite{Lowry1927,Landauer1978,Pier6,TorquatoBook,SahimiBookI,Tuncer2002a,Bergman1992,SihvolaBook,Tuncer2001a,dielectricfunction,BBReview,GunnarPhD,Tinga1973}. Mixture formulas based on analytical and effective medium approaches were developed, such that for various arrangement of inclusions predicting the dielectric properties of composites was plausible\cite{McPhedran2,mil81,Perrins1979,McPhedran1,mcp78}. A deep understanding of dielectric mixtures would be of great value (i) to be able to calculate either the dielectric constant of a mixture of substances of known dielectric constants or, (ii) knowing the dielectric constants of a mixture of two components and that of one of the components, to calculate the dielectric constant of the other~\cite{Lowry1927}, or (iii) even knowing the dielectric constants of a mixture of two components and that of two components to estimate the morphology of the mixture~\cite{TuncerSpectralPRB,TuncerJPD2005,Tuncer2005JPCMLET}. In late 1970's, Bergman cleverly showed that one can separate the geometrical contributions from the pure dielectric response of a composite if and only if the dielectric properties of the constituents were known \cite{bergman1982,Bergman3,Bergman1978}. Milton corrected errors in the Bergman's original derivation~\cite{Milton1981,Milton1981b,Miltona} and later Golden and Papanicolaou \cite{Golden1,Golden2} gave the rigorous derivation for the spectral representation theory. Recently, the present author has illustrated similarities between the dielectric relaxation and dielectric response of dielectric mixtures using the spectral density representation, the origin of similarities is very significant to comprehend physics of dielectrics \cite{Tuncer2005JPCMLET}.

The concept of having the cognition of the structure of composites, how the phases are arranged, is very useful in materials design, because special materials can be manufactured with the knowledge of structure-property relationship. For regular arrangement of phases, there exists equations based on theoretical calculations on simple enough geometries. However, fractal structures are abundant in nature \cite{Mandelbrot1982}, therefore to comprehend the materials properties with fractal structure has been a challenge for researchers for some decades. The fractal geometry or systems indicating hierarchical order has been one of the interesting topics in applied and theoretical (mathematical) physics \cite{Mandelbrot1977,Mandelbrot1982,AharonyBook,Fractal1989,Fractal1985}. As an example, electrical properties of metal aggregates in insulting matrix media were studied extensively \cite{GunnarFractals1989,gunnar1,gunnarPRB2002,Stroud1992,Stroud1986,GunnarSSC1986,Clerc1996,Clerc1994,Clerc1990}. In these studies either the dimension of the electrical network or the system, or the resonance frequency of the electrical impedance was used as a measure to indicate the fractal dimensions. When real systems are taken into account, the structural information is on the other hand usually obtained by optical/microscopic techniques, which are later analyzed to estimate the fractal dimensions. One can as well utilize a mixture formula to model the electrical properties of the composite system in hand, such that the model contain structural information, \eg\ there exist effective medium theories for composites with spherical and ellipsoidal inclusions \cite{Sillars1937,wiener,Pier6}. In the present paper, we employ the spectral density representation, which is a general representation for composites, to resolve the geometrical description of a model system described by the Landau-Lifshitz/Looyenga (LLL) effective medium formula \cite{LL,Looyenga}, or in other words we challenge the physical significance of the LLL expression. The LLL expression was used to describe the dielectric properties of dispersive systems composed of powders or exhibiting porous structure \cite{Spanier2000PRB,Marquardt1989,Pier6,NelsonPier,HaoDua2004,Kolokolova2001,BordiJNCS2002,BonincontroCSB1996,BordiCS1989,NelsonLLLMST,NelsonLLLJPD,NeeLLL,BenaddaLLL,DaviesLLL,LalLLL}. It was even showed \cite{Marquardt1989} that the LLL formula was more reliable when mixtures contained strongly dissipative particles and compared to others like Maxwell Garnett (MG) \cite{Levy1997,Maxwell_Garnett}, Bruggeman \cite{Bruggeman1935}, \etc\ (see for example Refs.~\cite{Lowry1927,Landauer1978,Pier6,TorquatoBook,SahimiBookI,Tuncer2002a,Bergman1992,SihvolaBook,Tuncer2001a,dielectricfunction,BBReview,GunnarPhD} for other formulas).

In this paper, we first present that the spectral density representation can in fact be written in a novel, more elegant, form that can be implemented in already existing dielectric data analysis techniques~\cite{Jonscher1983,RossEnisJP:CM,Tuncer2000b,MacDonald1987,LEVM}. Later we use the presented novel notation  and the numerical procedure to solve inverse problems \cite{TuncerLicDRT,TuncerSpectralPRB,TuncerLang,TuncerJPD2005} to test our hypothesis. We proceed to achieve our goal by considering the LLL~\cite{LL,Looyenga} expression for dielectric mixtures and discuss the significance of the presented approach on the mixture expression. 

The paper is organized as follows, first we present the spectral density representation for a binary mixture in \S \ref{SDR}, in this section the similarities between the dielectric relaxation in dielectrics and dielectric permittivity of binary mixtures are illustrated. \S \ref{Representation} describes the dielectric data representation and gives hints for analyzing impedance data of mixtures. The numerical method to solve the inverse integral is also presented explicitly for the interested readers in \S \ref{Numerical}. The numerical data generation and Landau-Lifshitz/Looyenga expression are presented in \S \ref{LLL}. The comparison of the results obtained by the inverse integral solution and the proposed conventional dielectric dispersion expression are given in \S \ref{havneg}. Conclusions are in \S \ref{conclusions}. 

\section{Spectral density representation}\label{SDR}
In the spectral density representation analysis of binary mixtures, the dielectric permittivity of a heterogeneous (effective) medium, is expressed as \cite{bergman1982,Bergman3,Bergman1978,Milton1981,Milton1981b,Miltona,Golden1,Golden2,GhoshFuchs,Gonc2000,Gonc2003},
  \begin{eqnarray}
    \label{eq:generalwiener_e}
    \vare=\varm\left\{1+q\left[A\left(\frac{\vari}{\varm}-1\right)+ {\int}_{0}^{1}\frac{{\sf G}(x) \dd x}{\left(\frac{\displaystyle \vari}{\displaystyle \varm}-1\right)^{-1}+x}\right]\right\}  
  \end{eqnarray}
where $\vare$, $\varm$ and $\vari$ are the complex dielectric permittivity of the effective, matrix and inclusion media, respectively; $q$ and $x$ are the concentration of inclusions and the spectral parameter, respectively. The function ${\sf G}(x)$ is the spectral density function (SDF), possesses information about the topological description of the mixture. Eq.~(\ref{eq:generalwiener_e}) can be arranged in a more elegant form, cf. \S~\ref{sec:appendix}, as follows,
\begin{eqnarray}
  \label{eq:spectral}
  \xi= \xi_s+q\int_0^1 \frac{{\sf G}(x) \dd x}{1+\varepsilon_{\sf m}^{-1}\Delta_{\sf im}x}
\end{eqnarray}
where, $\Delta_{\sf im}=\vari-\varm$, and $\xi$ is the complex and frequency dependent `scaled' permittivity, 
\begin{eqnarray}
  \label{eq:xi}
  \xi=\frac{\vare-\varm}{\vari-\varm},  
\end{eqnarray}
The constant $\xi_s$ in Eq.~\ref{eq:spectral} is complex and depends on the concentration and structure of the composite, its real part is related to the so called the `percolation strength'~\cite{GhoshFuchs,Day}. The mathematical properties and conditions that SDF satisfies are  presented in \S \ref{sec:appendix}~\cite{Bergman1978,GhoshFuchs,Stroud,Gonc2000,Gonc2003}. 


Eq.~(\ref{eq:xi}) is a very similar expression to the distribution of relaxation times (DRT) representation of a broad dielectric dispersion (relaxation) \cite{BottcherDRT,RossBJP,McCrum, Macdonald2000a,Macdonald2000b,macdonald:6241,TuncerLicDRT,Dias,Tuncer2004a,Tuncer2005JPCMLET},
\begin{eqnarray}
  \label{eq:drt}
  \ve(\imath\omega)= \ve_\infty+\Delta\ve\int_0^\infty \frac{{\sf G}(\tau) \dd \tau}{1+\imath\omega\tau}
\end{eqnarray}
where $\ve$, $\ve_\infty$ and $\Delta\ve$ are the complex permittivity, permittivity at optical frequencies and dielectric strength, respectively; and $\imath\equiv\sqrt{-1}$. The quantities $\omega$ and $\tau$ are the angular frequency and relaxation time, respectively. The distribution function for the relaxation times is $ {\sf G}(\tau)$. Comparison of Eq.~(\ref{eq:spectral}) and (\ref{eq:drt}) demonstrate that both the DRT and the scaled permittivity of SDF are actually the same. However, a new complex parameter $\varpi$ in SDF $\varpi\equiv\varm^{-1}\Delta_{\sf im}$ corresponds to the pure complex frequency $\imath\omega$ in DRT representation, and the real number constant $\ve_\infty$ is a complex number $\xi_s$ in SDF representation. In addition the spectral parameter $x$ corresponds to the relaxation time $\tau$ in the DRT. Finally, the dielectric strength $\Delta\ve$ in the DRT representation is related to the concentration of inclusions in the SDF. 

Due to the presented similarities or in other words the analogy, methods developed for dielectric data analysis\cite{Jonscher1983,MacDonald1987,LEVM} can be applied to the scaled permittivity $\xi$ of SDF \cite{TuncerJPD2005,TuncerSpectralPRB,Tuncer2005JPCMLET}. For example, one of the most employed dielectric dispersion expressions, known as Havriliak-Negami empirical expression \cite{HN}, can be used to analyze the scaled complex dielectric permittivity data of a mixture,
\begin{eqnarray}
  \label{eq:1}
  \xi(\varpi)=\xi_s+\frac{q}{[1+(\varpi x)^{\alpha}]^\beta}
\end{eqnarray}
where $\alpha$ and $\beta$ are parameters of a general distribution function \cite{HN,Tuncer2000b}, and $\varpi$ is the scaled complex frequency. When $\alpha=\beta=1$, a Debye-type relaxation is observed in the dielectric dispersion representation \cite{Debye1945}. In the case of spectral density representation, however, the Maxwell Garnett approximation is obtained for   $\alpha=\beta=1$ and $x=(1-q)/d$, since ${\sf G}(x)=\delta[x-(1-q)/d]$~\cite{Stroud,Gonc2000,Gonc2003,TuncerJPD2005,Tuncer2005JPCMLET,TuncerSpectralPRB}, where $d$ is the dimensionality of the system.

One can therefore in principle write a new more general empirical mixture formula~\cite{TuncerFormula} by isolating the dielectric permittivity $\vare$ of the composite in Eq.~(\ref{eq:xi}) and substituting it in Eq.~(\ref{eq:1}),
 \begin{eqnarray}
  \label{eq:empirical}
  \vare=\varm+\Delta_{\sf im}\left\{\xi_s+\frac{q}{[1+(\varpi x)^{\alpha}]^\beta}\right\}
\end{eqnarray}
Finally, note that the second, fractional expression inside the curly parenthesis in Eq.~(\ref{eq:empirical}) can be exchanged with any one of the dielectric dispersion relations existing in the literature~\cite{HN,CD,CC,Jonscher1983,MacDonald1987,NigmatullinJNCS2002,NigmatullinJPD2003,NigmatullinJPCM2003}.

\section{Representation of dielectric data}\label{Representation}
\begin{figure}[t]
  \centering
  \includegraphics[width=\figsize\linewidth]{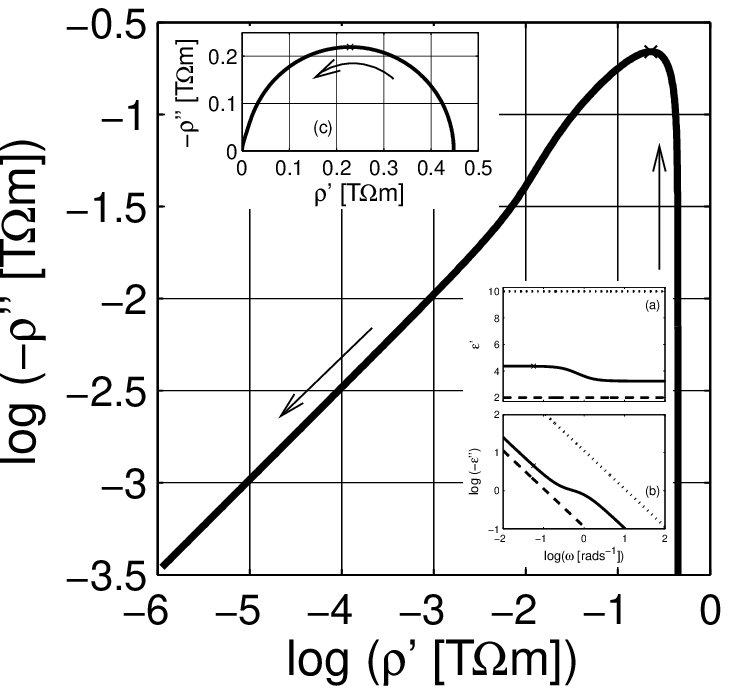}
  \caption{The Argand diagram of resistivity for a MG composite in log-log scale, the effective permittivity is calculated with $\vare=\varepsilon_{\sf e}^{\rm MG}(\omega; \varm,\,\vari,\,0.3,3)$, with $\varm=2+10^{-12}(\imath\varepsilon_0\omega)^{-1}$ and $\vari=10+10^{-10}(\imath\varepsilon_0\omega)^{-1}$. The arrows indicate the direction of increasing angular frequency $\omega$. The insets (a) and (b) are the real and the imaginary parts of permittivities; $\vare$, $\varm$ and $\vari$ are the permittivities for the matrix (\kesik), inclusion (\dotted) and effective (\full) media. The inset (c) is the Argand diagram of resistivity in linear-linear scale. Logarithmic scale is base 10.}
  \label{fig:RESMG}
\end{figure}
\begin{figure}[t]
  \centering
  \includegraphics[width=\figsize\linewidth]{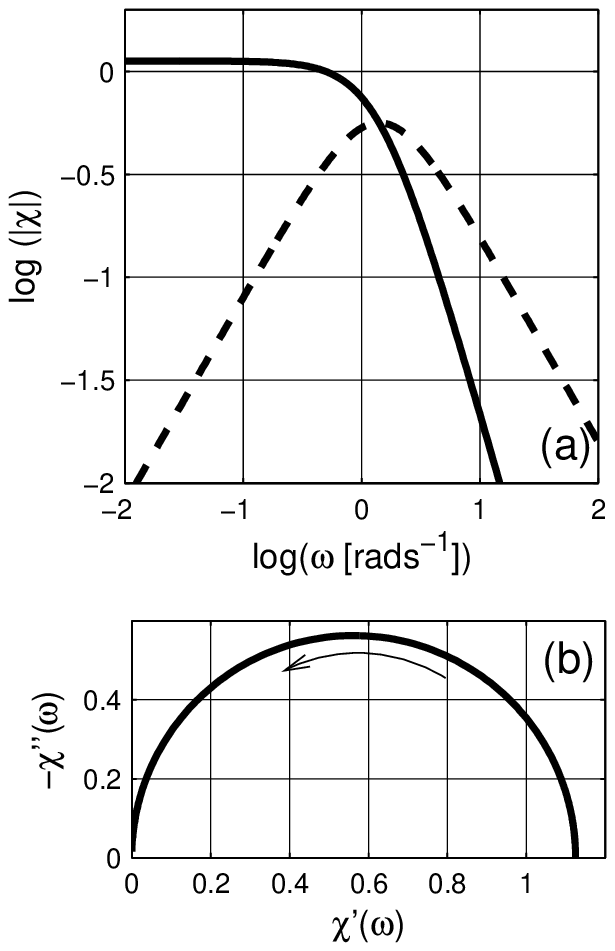}
  \caption{(a) Dielectric susceptibility $\chi$ as a function of angular frequency $\omega$; the real and the imaginary parts are presented with the solid (\full) and dashed (\kesik) lines, respectively. The arrow indicates the direction of increasing angular frequency $\omega$. (b) The Argand diagram (Cole-Cole plot) of susceptibility. Logarithmic scale is base 10.}
  \label{fig:CHIMG}
\end{figure}

The dielectric function for a $d$-dimensional (or composite with arbitrary shaped inclusions)  is defined as follows with Maxwell Garnett (MG) expression for a composite~\cite{Levy1997,Maxwell_Garnett} 
\begin{eqnarray}
  \label{eq:4}
  \varepsilon_{\sf e}^{\rm MG}(\omega;\, \varm,\,\vari,\,q,\,d)=\varm+ \frac{\varm\, d\, q\, \Delta_{\sf im}}{(1-q)\,\Delta_{\sf im} + d\, \varepsilon_{\sf m}}.
\end{eqnarray}
The dielectric data of the composite can be expressed in one of the four immittance representations\cite{RossEnisJP:CM,RossBJP,MacDonald1987},  ({\em i}) the complex resistivity $\rho(\omega)$; ({\em ii}) the complex modulus $M(\omega)\equiv\imath\omega\varepsilon_0\rho(\omega)$; ({\em iii}) the complex permittivity $\varepsilon\equiv[M(\omega)]^{-1}$; and ({\em iv}) the complex conductivity $\sigma(\omega)\equiv\imath\omega\varepsilon_0\varepsilon(\omega)\equiv[\rho(\omega)]^{-1}$. When we are dealing with frequency dependent dielectric properties of composites the effective conductivity of the composite can sometimes influence the imaginary part of the dielectric function--hinders the dielectric losses due to the interfacial polarization--as shown in inset (b) of Fig.~\ref{fig:RESMG}. In such cases it is more appropriate to use the complex resistivity representation (plot) as shown in the Argand diagrams in log-log and linear scales in Fig.~\ref{fig:RESMG} and \ref{fig:RESMG}c, respectively. The ohmic conductivity $\sigma_{{\sf e},dc}$ (or resistivity  $\rho_{{\sf e},dc}$) of the material can be estimated from the complex resistivity Argand plot. However once the conductivity contributions are  cleared from the immittance or dielectric data, using the estimated resistivity value in Fig.~\ref{fig:RESMG} as $\omega\rightarrow0$, the pure dielectric dispersion (permittivity) would be obtained. In addition the high frequency dielectric permittivity $\varepsilon(\omega\rightarrow\infty)\equiv\varepsilon_\infty$ can be further subtracted from the data to obtain the pure dielectric polarization (susceptibility $\chi$) of the composite as presented in Fig.~\ref{fig:CHIMG}; $\chi=\chi'-\imath\chi''\equiv\varepsilon-\varepsilon_{\infty}+\imath\sigma_{dc}(\varepsilon_0 \omega)^{-1}$. The imaginary part of $\chi$ has a peak around $\omega\sim1^{+}\ \rad\reciprocal\second$. The linear scale plot of the susceptibility $\chi$ is a semi circular curve, cf. Fig.~\ref{fig:CHIMG}b, as seen in Fig.~\ref{fig:RESMG}c.

\begin{figure}[t]
  \centering
  \includegraphics[width=\figsize\linewidth]{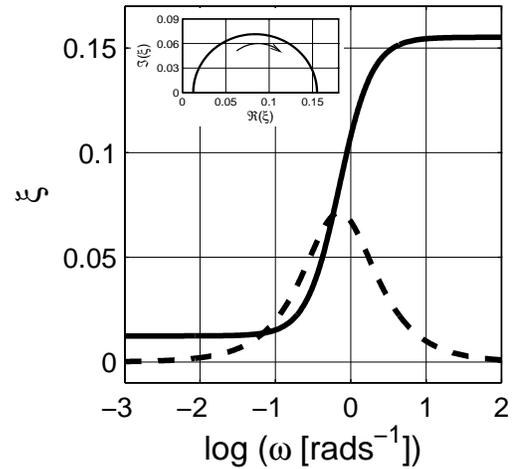}
  \caption{The real ($\Re(\xi)$; \full) and the imaginary ($\Im(\xi)$; \kesik) parts of scaled permittivity $\xi$ in the Maxwell Garnett approximation. The inset is the Argand diagram for $\xi$, the arrow shows the direction for increasing angular frequency $\omega$.  The inclusions are spherical $d=3$, and $q=0.3$. Logarithmic scale is base 10.}
  \label{fig:XIXIMG}
\end{figure}

\begin{figure}[t]
  \centering
  \includegraphics[width=\figsize\linewidth]{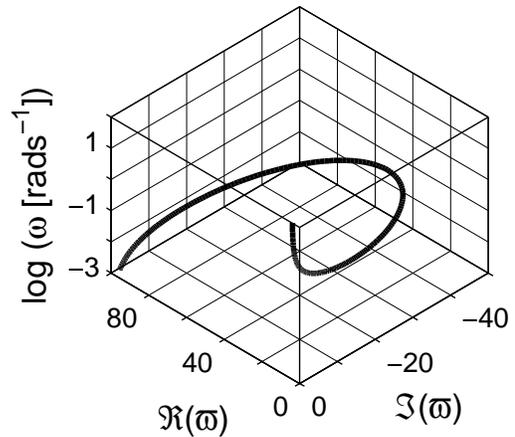}
  \caption{Dependence of the real frequency $\omega$ as a function of complex `scaled' frequency $\varpi$ in the spectral representation. Logarithmic scale is base 10.}
  \label{fig:OME3DMG}
\end{figure}

\begin{figure*}[t]
  \centering
  \includegraphics[width=\figsize\linewidth]{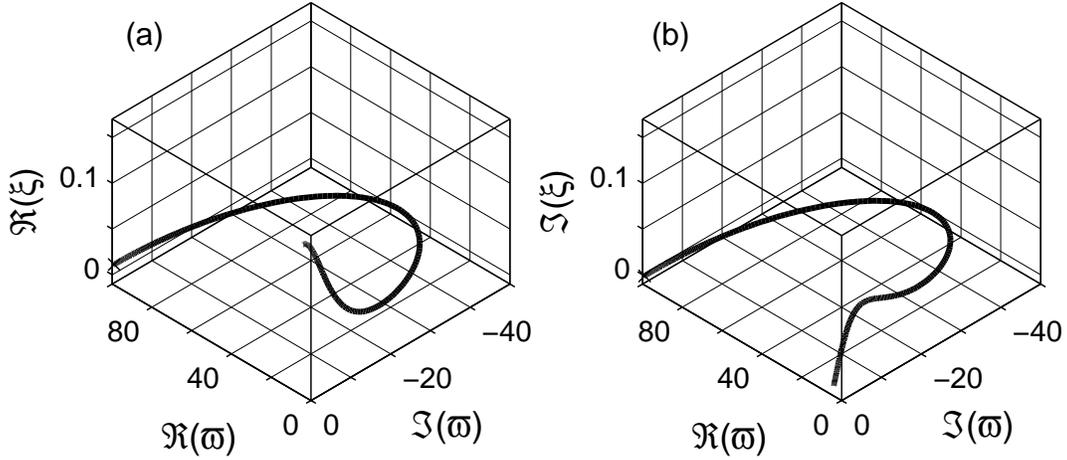}
  \caption{Three dimensional line plots of (a) the real $\Re(\xi)$ and (b) the imaginary $\Im(\xi)$ parts of the `scaled' permittivity for a Maxwell Garnett mixture. The inclusions are spherical $d=3$, and $q=0.3$.}
  \label{fig:SDR3DMG}
\end{figure*}

If we now consider the frequency dependent properties of the scaled permittivity $\xi$ for the considered MG composite, the real part and the imaginary parts are similar to the dielectric permittivity $\vare$ of the composite, cf. inset in Fig.~\ref{fig:XIXIMG}, which shows the  Argand diagram of the scaled permittivity $\xi$, observe that the increasing frequency is in the opposite direction when compared to the Argand plot of the susceptibility in Fig.~\ref{fig:CHIMG}b. The real part of $\xi$ is a mirror image of $\Re(\vare)$. Unlike the imaginary part of $\vare$ (due to the ohmic losses), $\Im(\xi)$ shows a clear peak around $\omega\sim1^{-}\ \rad\reciprocal\second$. The shift in the origin position in the Argand diagram in the inset of Fig.~\ref{fig:XIXIMG} is related to the percolation strength $\xi_s$, which is close to zero, and the concentration of the inclusions. There are similarities between the susceptibility and the scaled permittivity plots when the same frequency $\omega$ axis is used, however, note that the scaled frequency $\varpi$ for the scaled permittivity is a complex quantity. We therefore illustrate the dependence of the real angular frequency $\omega$ as a function of $\varpi$ in a 3D curve-plot in Fig.~\ref{fig:OME3DMG}. In addition the real and the imaginary parts of the scaled permittivity $\xi$ are shown in Fig.~\ref{fig:SDR3DMG} as a function of $\varpi$. As shown in the figure the actual dependence of $\xi$ on $\varpi$ is more complicated than $\varepsilon$ on $\omega$. On the contrary, this dependence can be used to estimate the SDF for a given system, which is explicitly given with a numerical procedure in the next section. 
  

\section{Numerical estimation of spectral density function}\label{Numerical}

The derived spectral density expression in Eq.~\ref{eq:spectral} is a Bolter equation~\cite{Volterra}, which is a special form of the Fredholm integral equations~\cite{Fredholm}. Such equations are usually considered to be {\em ill-conditioned} because of their non-unique solutions. However, the approach used here and recently presented elsewhere~\cite{Tuncer2004a,Tuncer2000b,TuncerSpectralPRB,TuncerLicDRT} leads to unique solutions by means of a constrained least-squares fit and the Monte Carlo integration methods. Some other approaches were also suggested to solve the spectral density function in the literature \cite{DayPhysB1,Day,DayPRL2000,GhoshFuchs,Gonc2003,Gonc2000,Cherkaev2003,Stroud1999,Stroud}. The presented numerical method to solve inverse integral transforms has previously been used in different problems~\cite{RossEnisJP:CM,Tuncer2004a,TuncerSpectralPRB,Tuncer2000b,TuncerLicDRT,TuncerLang,TuncerJPD2005}. In this particular approach, the integral in Eq.~(\ref{eq:xi}) is first written in a summation form over some number of randomly selected and fixed $x_n$-values, $x_n\in [0,1]$, where $n$ is less than the total number $M$ of experimental (known) data points in the complex scaled permittivity, $\xi$, 
\begin{eqnarray}
  \label{eq:summation}
  \xi= \xi_s+\sum \frac{{\sf g}_{n}}{1+\varpi x_n}\quad n\le M
\end{eqnarray}
This  converts the non-linear problem in hand to a linear one with ${\sf g}_{n}$ being the unknowns, weights of the randomly selected $x_n$ values. In the present notation 
${\sf g}$ is $q\,{\sf G}$.  
Later, a constrained least-squares algorithm is applied to get the corresponding ${\sf g}$-values and $\xi_s$,
\begin{eqnarray}
  \label{eq:2}
  \min\sum\left[\underline{\xi}-\mathfrak{K}\mathfrak{g}\right]^2 \quad {\rm and} \quad \mathfrak{g}\ge 0
\end{eqnarray}
where $\mathfrak{K}$ is the kernel-matrix, 
\begin{eqnarray}
  \label{eq:kernelmatrix}
  \mathfrak{K}=\left( 
    \begin{array}{cccc}
      1 & {\bf K}_{11} & {\bf K}_{12} & \ldots\\
      1 & {\bf K}_{21} & {\bf K}_{22} & \ldots\\
      1 & {\bf K}_{31} & {\bf K}_{32} & \ldots\\
      \vdots &      \vdots &      \vdots &      \ddots \\
    \end{array}
\right).
\end{eqnarray}
Here ${\bf K}_{ij}=[1+\varpi_i x_j]^{-1}$, index $i$ runs on the angular frequency points $i=1,\,\dots,\,M$, and index $j$ runs on the randomly selected $x$ values, $j=1,\,\dots,\,n$.  The parameters $\underline{\xi}$ and $\mathfrak{g}$ in Eq.~(\ref{eq:2}) are column vectors, respectively, the scaled permittivity calculated from the experimental (known) data and the searched spectral density, 
\begin{eqnarray}
  \label{eq:gvector}
 \underline{\xi}=\left( 
    \begin{array}{c}
      \xi[\varpi(\omega_1)]\\
      \xi[\varpi(\omega_2)]\\
      \xi[\varpi(\omega_3)]\\
      \xi[\varpi(\omega_4)]\\
      \vdots  \\
    \end{array}
\right)&\quad \text{and} \quad&
 \mathfrak{g}=\left( 
    \begin{array}{c}
      \xi_s\\
      {\sf g}_1\\
      {\sf g}_2\\
      {\sf g}_3\\
      \vdots  \\
    \end{array}
\right)  
\end{eqnarray}
In our numerical procedure, we perform many minimization steps with fresh, new,  sets of randomly selected $x_j$-values. The ${\sf g}_{j}$-values and $\xi_s$ obtained are recorded in each step, which later build-up the spectral density distribution ${\sf g}$ and a distribution for the percolation strength, $\xi_s$. For a large number of minimization loops, actually the $x$-axis becomes continuous---the Monte Carlo integration hypothesis---contrary to regularization methods~\cite{Day,Cherkaev2003,macdonald:6241}. In the presented analysis below, the total number of randomly selected $x$ values are $\sim10^6$. The number $M$ of data points are chosen 24, and the number $N$ of unknown ${\sf g}$-values are 22. 

Application of the numerical procedure to the Maxwell Garnett expression, impedance data of porous rock-brine mixture and two-dimensional `ideal' structures has previously been presented elsewhere\cite{TuncerSpectralPRB,TuncerJPD2005}. The estimated spectral density functions for the MG expression were delta sequences~\cite{Butkov} as expected, without any significant percolation component, because of the estimated concentration $q$, cf. Eq.~(\ref{eq:A51}). In the next section, we apply the numerical procedure to the Landau-Lifshitz/Looyenga~\cite{LL,Looyenga} expression in order to better understand the nature of dielectric mixtures, which obey this relation.

\section{Landau-Lifshitz/Looyenga expression}\label{LLL}

\begin{figure}[t]
  \centering
  \includegraphics[width=\figsize\linewidth]{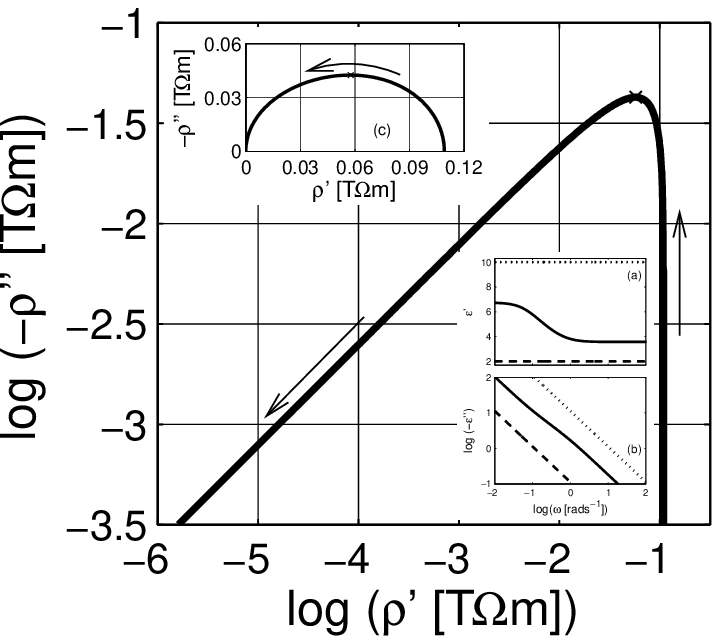}
  \caption{The Argand diagram of resistivity for a LLL composite in log-log scale, the effective permittivity is calculated by $\vare=\vare^{\rm LLL}(\omega; \varm,\,\vari,\,0.3,\,3)$ with $\varm=2+10^{-12}(\imath\varepsilon_0\omega)^{-1}$ and $\vari=10+10^{-10}(\imath\varepsilon_0\omega)^{-1}$. The arrows indicate the direction of increasing angular frequency $\omega$. The insets (a) and (b) are the real and the imaginary parts of permittivities; $\vare$, $\varm$ and $\vari$ are the permittivities for the matrix (\kesik), inclusion (\dotted) and effective (\full) media. The inset (c) is the Argand diagram of resistivity in linear-linear scale. Logarithmic scale is base 10.}
  \label{fig:RESLLL}
\end{figure}
\begin{figure}[t]
  \centering
  \includegraphics[width=\figsize\linewidth]{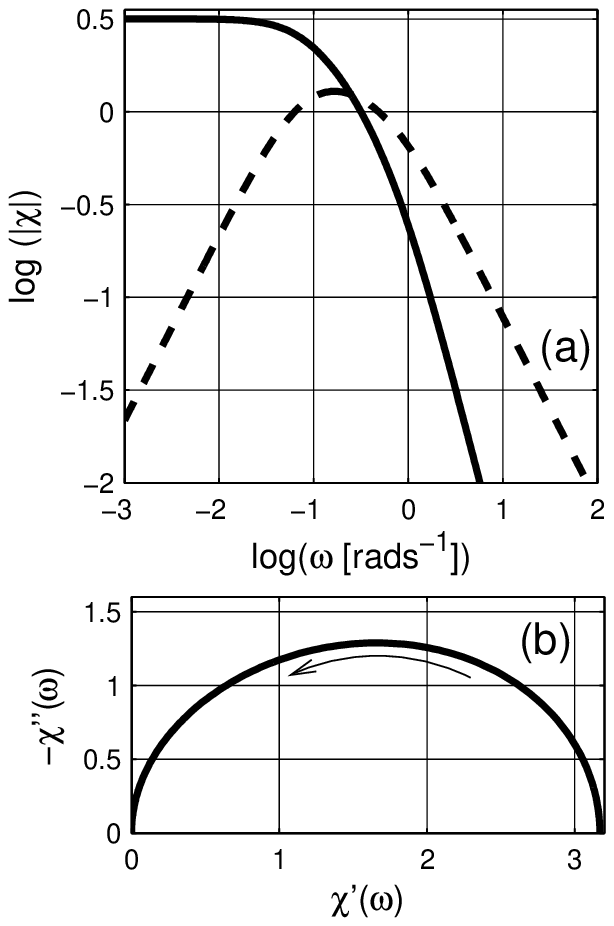}
  \caption{(a) Dielectric susceptibility $\chi$ as a function of angular frequency $\omega$; the real and the imaginary parts are presented with the solid (\full) and dashed (\kesik) lines, respectively. The arrow indicates the direction of increasing angular frequency $\omega$. (b) The Argand diagram (Cole-Cole plot) of susceptibility. Logarithmic scale is base 10.}
  \label{fig:CHILLL}
\end{figure}

\begin{figure}[t]
  \centering
  \includegraphics[width=\figsize\linewidth]{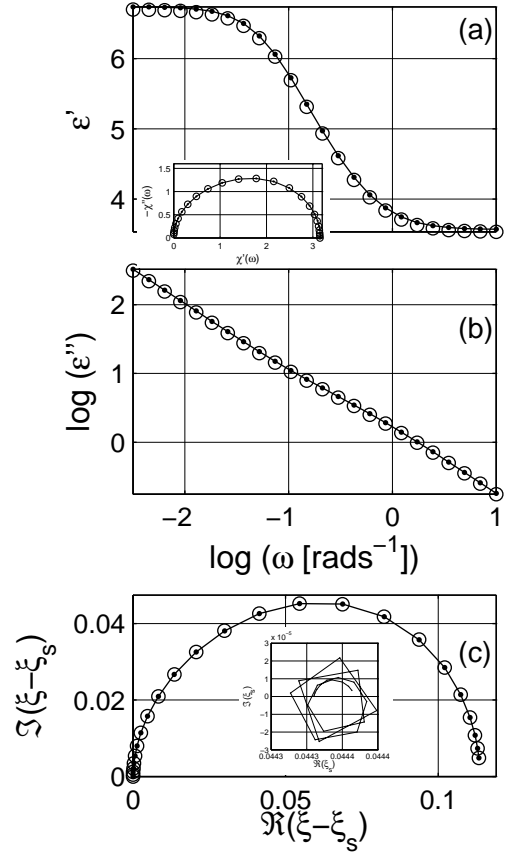}
  \caption{(a) The real and (b) the imaginary parts of the dielectric permittivity calculated with Eq.~(\ref{eq:LLL}) for $q=0.3$, the permittivities of the matrix and the inclusion phases are $\varm=2+10^{-12}(\imath\varepsilon_0\omega)^{-1}$ and $\vari=10+10^{-10}(\imath\varepsilon_0\omega)^{-1}$. (c) Argand diagram of scaled permittivity without out the percolation strength contribution.  The inset in (a) shows the dielectric susceptibility after the subtraction of ohmic conductivity and the permittivity at high frequencies. The inset in (c) shows the Argand plot of $\xi_s$ for each point, the values are very narrowly distributed. The lines with points are the simulated data of Eq.~\eqref{eq:LLL} and the symbols ($\circ$) are the dielectric response calculated with the estimated spectral density function from the proposed numerical algorithm. There is a very good agreement between simulated and analyzed data sets.  Logarithmic scale is base 10.}
  \label{fig:dataLLL3}
\end{figure}

Landau and Lifshitz \cite{LL} and Looyenga \cite{Looyenga} independently, using different approaches developed an expression for dielectric mixtures, implies the additivity of cube roots of the permittivities of the mixture constituents when taken in proportion to their volume fractions (see Refs. \cite{Spanier2000PRB,Marquardt1989,Pier6,NelsonPier,HaoDua2004,Kolokolova2001,BordiJNCS2002,BonincontroCSB1996,BordiCS1989,NelsonLLLMST,NelsonLLLJPD,NeeLLL,BenaddaLLL,DaviesLLL,LalLLL} for examples).
\begin{eqnarray}
  \label{eq:LLL}
  [\vare^{\rm LLL}(\omega;\varm,\,\vari,\,q)]^{1/3}=(1-q)\,\varm^{1/3}+q\,\vari^{1/3}
\end{eqnarray}
This expression is extensively used in the literature for powdered materials and optical properties of material mixtures \cite{NelsonPier,Pier6}. In the following calculations, we choose the same values for the dielectric functions of the phases as before, cf. Fig.~\ref{fig:RESMG}. 
The concentration $q$ of inclusion phase is varied between 0.1  and 0.9 in the simulations.

\begin{figure}[t]
  \centering
  \includegraphics[width=\figsize\linewidth]{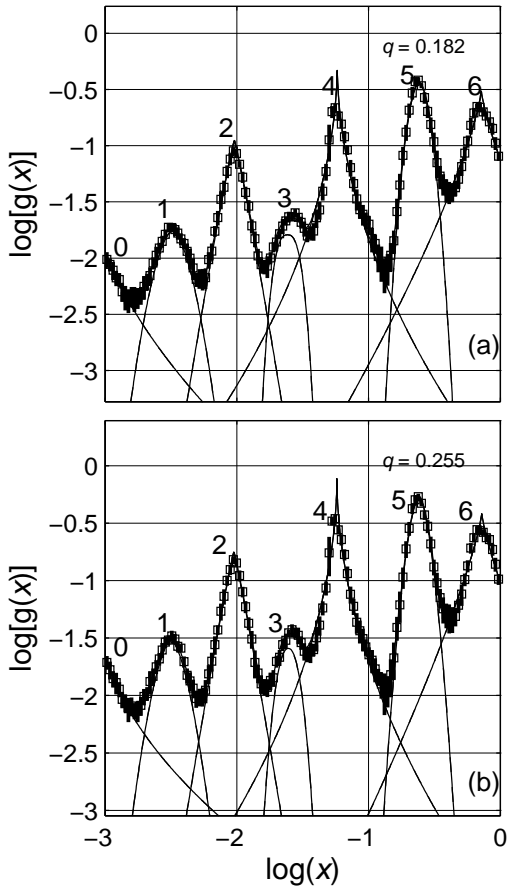}
  \caption{Spectral density functions of Landau-Lifshitz/Looyenga equation for two concentrations; (a) $q=0.2$ and (b) $q=0.3$. The solid lines (\full) are L{\'e}vy distributions adapted to the estimated peaks. The numerical data is presented with symbol, error bars. The numbers on the peaks denote the significant peaks. Logarithmic scale is base 10.}
  \label{fig:Levyfit}
\end{figure}

\begin{figure}[t]
  \centering
  \includegraphics[width=\figsize\linewidth]{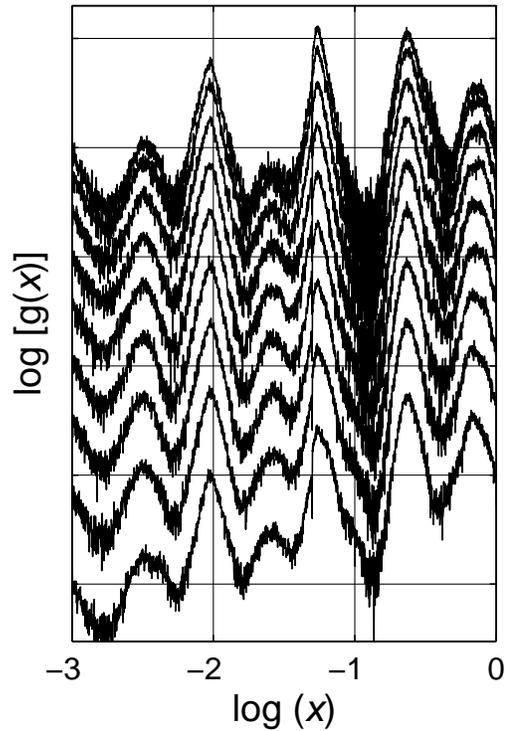}
  \caption{Spectral density functions of Landau-Lifshitz/Looyenga equation for nine different concentrations, $q=\{0.1,\,\dots,\,0.9\}$. The data are shifted for better comparison. There are six peaks resolved between $x=10^{-3}$ and $x=1$. Interestingly, the positions of the most probable spectral parameters of the six are not altered with increasing concentration. Logarithmic scale is base 10.}
  \label{fig:LLL_g}
\end{figure}

The extracted spectral functions are like distributions, and they are analyzed by means of comparing them with a known distribution. We apply the L{\'e}vy statistics~\cite{BreimanLevy,LoeveBook,Walter1999,Donth2002}, which is widely used for interacting systems in different research fields~\cite{Levy,Barkai2000,Barkai2002,Furukawa1993,Stoneham1969,Walter1999,Donth2002,TuncerSpectralPRB,Tuncer2004a}. The L{\'e}vy stable distribution is a natural generalization (approximation) of the normal (Gaussian), Cauchy or Lorenz and Gamma distributions. It is used when analyzing sums of independent identically distributed random variables by a diverging variance. 
Its characteristic function is expressed as
\begin{eqnarray}
  \label{eq:Levy}
  {\sf L}(x;\mathfrak{A},\mu,\gamma,\zeta)=\mathfrak{A}|\exp\{ - |\zeta(x-\mu)|^\gamma\}|
\end{eqnarray}
Here,  $\gamma$ is the characteristic exponent ($\gamma>0$), $\mu$ is the localization parameter, $\zeta$ is the scale parameter and $\mathfrak{A}$ is the amplitude. The special forms of Eq.~(\ref{eq:Levy}) are the Gaussian [${\sf L}(x;\,A,\,\mu,\,2,\,\zeta)$], the Lorentz or Cauchy [${\sf L}(x;\,A,\,\mu,\,1,\,\zeta)$] and Gamma [${\sf L}(x;\,A,\,\mu,\,1/2,\,\zeta)$] distributions. Different forms of probability density functions for L{\'e}vy statistics exists, we have adopted a stable distribution used in the literature \cite{BreimanLevy,LoeveBook,Walter1999}. We omit the imaginary parts in the characteristic function because of their insignificance in the results. 

In Fig.~\ref{fig:RESLLL}, the simulated dielectric permittivity with Eq.~(\ref{eq:LLL}) is presented in the complex resistivity level. The actual dielectric data is shown in the insets Fig.~\ref{fig:RESLLL}a and \ref{fig:RESLLL}b. Compared to the MG expression in Eq.~\eqref{eq:4}, shown in Fig.~\ref{fig:RESMG}, the LLL expression does not show a knee-point as the complex resistivity decreases with increasing frequency. In addition the linear scale Argand diagram of the LLL expression in Fig.~\ref{fig:RESLLL}c is not a perfect semi circle as the MG one. The dielectric susceptibility after the subtraction of the frequency independent parameters $\varepsilon_\infty$ and $\sigma$ are presented in Fig.~\ref{fig:CHILLL}. The losses, $\chi''$, are non-symmetrical for the LLL expression, which indicates that the actual dielectric response can be modeled by a non-Debye dielectric dispersion, \eg\ the Havriliak-Negami expression, this is also visible in the Argand diagram of the susceptibility, Fig.~\ref{fig:CHILLL}b.

The reconstructed complex dielectric permittivity and the scaled permittivity are presented in Fig.~\ref{fig:dataLLL3}. 
It is important to mention that the fitting is performed in the scaled permittivity level. While numerically calculating the spectral function the randomly selected spectral parameters $x$ are picked between $10^{-3}$ and $1$ in logarithmic scale; the pre-distribution is log-linear (for details see \cite{TuncerLicDRT}). In such an integration limit in Eq.~(\ref{eq:spectral}), the distribution or the functional contributions of $x$-values lower than $10^{-3}$ are included in the percolation strength $\xi_s$. 

\begin{figure}[t]
  \centering
  \includegraphics[width=\figsize\linewidth]{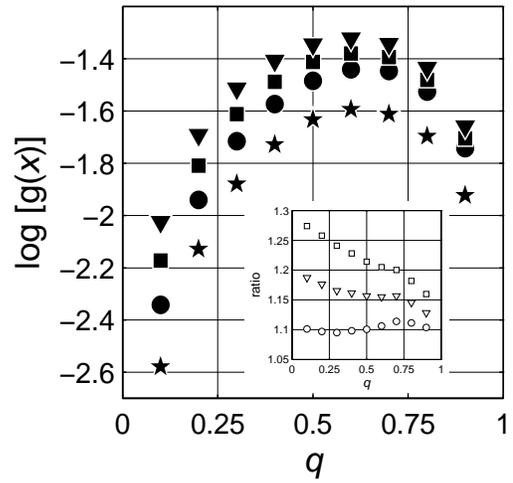}
  \caption{Spectral density function values at constant $x$, [$\log(x)=-2.5$ ($\bigstar$); $\log(x)=-2$ ($\bullet$); $\log(x)=-1.275$ ($\blacktriangledown$) and $\log(x)=-0.675$ ($\blacksquare$)]. The inset illustrates the ratios for $\log \{{\sf g}[\log(x)=-2.5]\}/\log \{{\sf g}[\log(x)=-2]\}$ ($\Box$), $\log \{{\sf g}[\log(x)=-2.5]\}/\log \{{\sf g}[\log(x)=-1.275]\}$ ($\triangledown$) and $\log \{{\sf g}[\log(x)=-2.5]\}/\log \{{\sf g}[\log(x)=-0.675]\}$ ($\circ$).  Logarithmic scale is base 10.}
  \label{fig:atcontg}
\end{figure}

\begin{figure}[t]
  \centering
  \includegraphics[width=\figsize\linewidth]{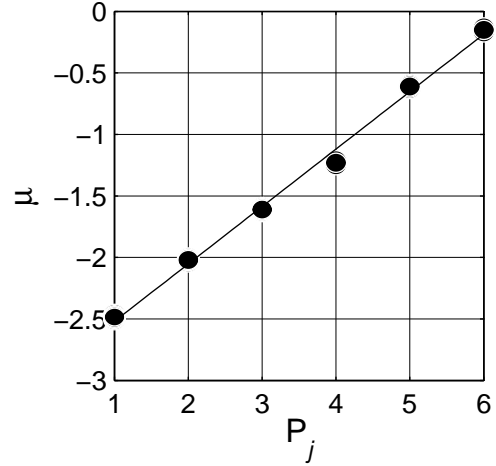}
  \caption{Peak positions, $\mu=\log(\overline{x})$. The solid line is a linear fit with $\mu=0.470P_j-2.99$, where $j$ is the peak number in Table~\ref{tableSDR}.}  \label{fig:fractal}
\end{figure}

The estimated spectral density functions for two concentration levels, $q=0.2$ and $q=0.3$, are presented in Fig.~\ref{fig:Levyfit}. There are six visible peaks, which are labeled with numbers one to six from left to right, respectively. We do not take into account peak 0 (zero) in the analysis. The estimated integral of the distributions are presented on the graphs. Each peak is analyzed by the L{\'e}vy distribution as mentioned before, and the solid lines (\full) show the individual distributions, cf. Fig.~\ref{fig:Levyfit}. It is remarkable that the estimated distributions are similar in form but shifted a little bit up with increase in concentration $q$. This behavior is an illustration of self-similar  fractal nature of the considered composite system in the LLL expression, in which the topological arrangement is not changing significantly with increased concentration. To support this argument/statement, the spectral density functions of mixtures described with the LLL expression for nine different concentration are shown in Fig.~\ref{fig:LLL_g}. The shifting is performed with constant steps in concentration for clarity, however as observed the spectral function amplitudes are not increasing proportional with increasing inclusion concentration, cf. Table~\ref{tableSDR} and cf. Fig.~\ref{fig:atcontg}. The difference between peak positions is constant in logarithmic scale $|P_j-(P_{j-1}|\approx0.47$, cf. Fig.~\ref{fig:fractal}.

In Table~\ref{tableSDR} not only the positions but also the shape parameters of the L{\'e}vy expression are listed for the six peaks resolved for each concentration.  The peak positions $\mu$ are not varied with increasing concentration of the inclusions phase. The amplitude $\mathfrak{A}$ changes with increasing concentration for each peak. The scale parameter $\zeta$ and the characteristic exponent $\gamma$ indicate some relation to concentration, the their exact relations are not sought. In Fig.~\ref{fig:atcontg}, we show the change in the amplitude for four of the peaks with increase in the concentration of the inclusions.  It is remarkable that around $q=0.6$ the behavior of the spectral density functions change, the increase in the amplitude of the peaks with increasing concentration starts to decrease for increasing concentration as $q>0.6$. In the inset the ratio of the three selected peaks to peak 1 are shown in the figure, a similar activity is also observed in the ratio. Since the ratio between the amplitudes of the selected peaks does not indicate a simple linear relation to concentration $q$, the topological description of the system can not be qualitatively investigated. However, as mentioned previously the location and form of the spectral density functions indicate a huge resemblance to each other that they are in fact related to the self-similar hierarchical nature of the composites expressed by LLL expression.

\begin{figure}[t]
  \centering
  \includegraphics[width=\figsize\linewidth]{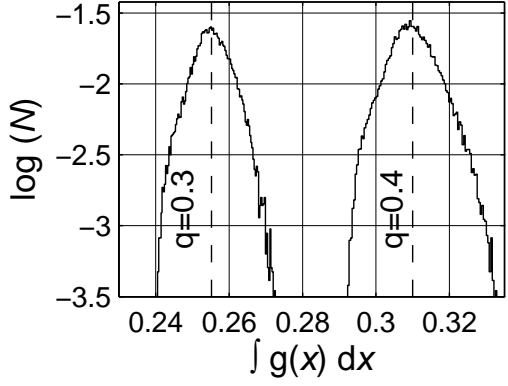}
  \caption{The number density distribution of estimated $\int {\sf g}(x) \dd x\equiv\overline{q}$ for two concentration $q$ levels, $q=\{0.3,\,0.4\}$. The dashed vertical lines are the expectation values of $\overline{q}$, which are $0.255\pm0.065$ and $0.310\pm0.096$ for concentration levels $0.3$ and $0.4$, respectively.   Logarithmic scale is base 10.}
  \label{fig:LLL_histg}
\end{figure}

\begin{figure}[t]
  \centering
  \includegraphics[width=\figsize\linewidth]{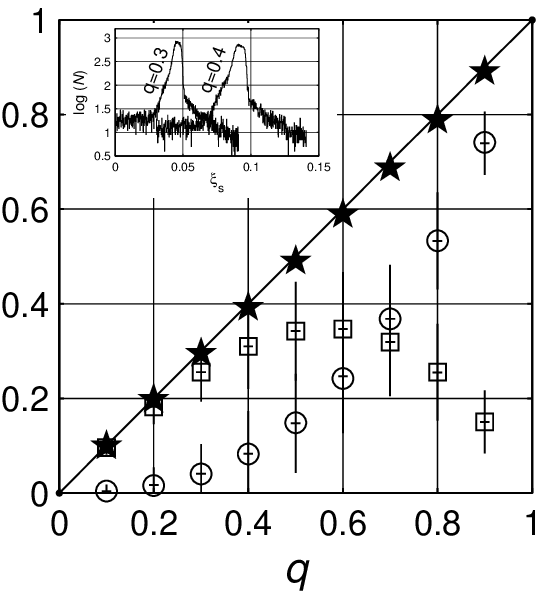}
  \caption{The expectation values of constants, $\overline{q}$ and $\overline{\xi_s}$, in the spectral density approach. The integral of ${\sf g}(x)$ yields the concentration of the inclusions $\overline{q}$, denoted by ($\Box$) symbols. The expected value of percolation strength is $\overline{\xi}$ denoted by ($\circ$) symbols. Both values are also presented with error bars. Addition of the two expected values $\overline{\xi_s}+\overline{q}$, denoted with ($\bigstar$) symbols, leads to the actual concentration $q$ as in Eq.~(\ref{eq:A51}), which is denoted by the solid line (\full). In the inset, the number distributions of $\xi_s$, percolation strenght, estimated in each Monte Carlo cycle for $q=0.3$ and $q=0.4$ are illustrated, respectively, the most expected $\overline{\xi_s}$ are $0.041\pm0.006$ and $0.083\pm0.015$ for the two concentrations.}
  \label{fig:LLL_Aandg}
\end{figure}

Finally the statistical analysis of the concentration calculated from the integral of spectral density function [inregral expression on the right-hand side of Eq.~\eqref{eq:spectral}], and the value of the percolation strength $\xi_s$ at each Monte Carlo step are presented as summarized in Fig.~\ref{fig:LLL_histg} and \ref{fig:LLL_Aandg}. The number distribution of integral of spectral density function calculated at each Monte Carlo cycle is not centered at the actual concentration $q$ taken but deviated. The deviation indicates that there is a percolation path, network structure, cf. Fig.~\ref{fig:LLL_histg}. As expected from the definition of the spectral density function, Eq.~(\ref{eq:A51}), addition of the estimated concentration $\overline{q}$ and percolation strength $\overline{\xi_s}$, presented with error bars and open symbols ($\Box$) and ($\bigcirc$), respectively, yield very close numerical values as the actual concentration $q$. 
The value $\overline{q}$ is calculated from the distribution. It is striking that the applied numerical method is capable of estimating the concentration of the filler material when it is not known in advance. 
\begin{table*}
\caption{Fit parameters of L{\'e}vy distribution, Eq.~\ref{eq:Levy}, for the seven peaks estimated by the numerical algorithm.\label{tableSDR}}
\begin{tabular*}{\linewidth}{c@{\extracolsep{\fill}}rrrrr||crrrrr}
\hline\hline
Peak & $q$   &  $\mathfrak{A}$  & $\mu$ &$\gamma$& $\zeta$ & Peak & $q$   &  $\mathfrak{A}$  & $\mu$ &$\gamma$ & $\zeta$ \\ 
\hline\hline
1 & 0.1 &  0.01 & -2.47 &  1.58 &  6.18 &2 & 0.1 &  0.06 & -2.01 &  0.89 & 15.84\\
 & 0.2 &  0.02 & -2.49 &  1.71 &  6.57 & & 0.2 &  0.11 & -2.02 &  1.10 & 12.74 \\
 & 0.3 &  0.03 & -2.50 &  1.66 &  7.29 & & 0.3 &  0.18 & -2.02 &  1.09 & 12.83 \\
 & 0.4 &  0.04 & -2.50 &  1.75 &  7.83 & & 0.4 &  0.22 & -2.02 &  1.20 & 11.81 \\
 & 0.5 &  0.04 & -2.50 &  1.94 &  7.53 & & 0.5 &  0.26 & -2.03 &  1.30 & 11.43 \\
 & 0.6 &  0.04 & -2.49 &  2.06 &  7.81 & & 0.6 &  0.28 & -2.03 &  1.33 & 11.46 \\
 & 0.7 &  0.04 & -2.49 &  1.98 &  7.86 & & 0.7 &  0.28 & -2.03 &  1.22 & 11.84 \\
 & 0.8 &  0.04 & -2.48 &  2.09 &  7.59 & & 0.8 &  0.24 & -2.03 &  1.26 & 11.63 \\
 & 0.9 &  0.03 & -2.48 &  1.79 &  7.20 & & 0.9 &  0.16 & -2.02 &  1.15 & 11.54 \\
\hline
3 & 0.1 &  0.01 & -1.62 &  2.97 &  8.23&4 & 0.1 &  0.23 & -1.22 &  0.58 & 35.98 \\
 & 0.2 &  0.02 & -1.62 &  2.70 &  8.26 &  & 0.2 &  0.54 & -1.24 &  0.52 & 49.50\\
 & 0.3 &  0.03 & -1.61 &  2.21 &  9.34 &  & 0.3 &  0.82 & -1.24 &  0.52 & 53.04\\
 & 0.4 &  0.03 & -1.61 &  2.27 &  9.10 &  & 0.4 &  1.04 & -1.24 &  0.56 & 49.16\\
 & 0.5 &  0.04 & -1.61 &  2.51 &  8.84 &  & 0.5 &  1.20 & -1.24 &  0.57 & 48.10\\
 & 0.6 &  0.04 & -1.60 &  2.22 &  9.05 &  & 0.6 &  1.30 & -1.24 &  0.60 & 47.57\\
 & 0.7 &  0.03 & -1.61 &  2.58 &  8.91 &  & 0.7 &  1.26 & -1.24 &  0.60 & 48.15\\
 & 0.8 &  0.02 & -1.61 &  2.34 &  8.69 &  & 0.8 &  1.05 & -1.24 &  0.62 & 45.98\\
 & 0.9 &  0.01 & -1.62 &  2.43 &  8.32 &  & 0.9 &  0.58 & -1.24 &  0.62 & 41.94\\
\hline
5 & 0.1 &  0.20 & -0.62 &  1.85 & 10.89 & 6 & 0.1 &  0.22 & -0.16 &  0.67 & 15.10\\
 & 0.2 &  0.36 & -0.62 &  1.90 & 10.51 & & 0.2 &  0.31 & -0.14 &  0.77 & 10.98 \\
 & 0.3 &  0.54 & -0.62 &  1.70 & 11.15 & & 0.3 &  0.38 & -0.14 &  0.81 & 10.58 \\
 & 0.4 &  0.62 & -0.62 &  1.72 & 10.89 & & 0.4 &  0.40 & -0.14 &  0.80 &  9.84 \\
 & 0.5 &  0.69 & -0.61 &  1.58 & 10.90 & & 0.5 &  0.37 & -0.14 &  0.97 &  9.08 \\
 & 0.6 &  0.65 & -0.61 &  1.63 & 10.25 & & 0.6 &  0.33 & -0.14 &  1.02 &  8.87 \\
 & 0.7 &  0.56 & -0.61 &  1.66 &  9.57 & & 0.7 &  0.28 & -0.14 &  1.08 &  8.93 \\
 & 0.8 &  0.39 & -0.60 &  1.80 &  8.46 & & 0.8 &  0.18 & -0.14 &  1.33 &  8.22 \\
 & 0.9 &  0.26 & -0.60 &  1.50 &  9.33 & & 0.9 &  0.10 & -0.14 &  1.34 &  8.45 \\
\hline\hline
\end{tabular*}
\end{table*}

\section{Application of the Havriliak-Negami expression}\label{havneg}

\begin{figure}[t]
  \centering
  \includegraphics[width=\figsize\linewidth]{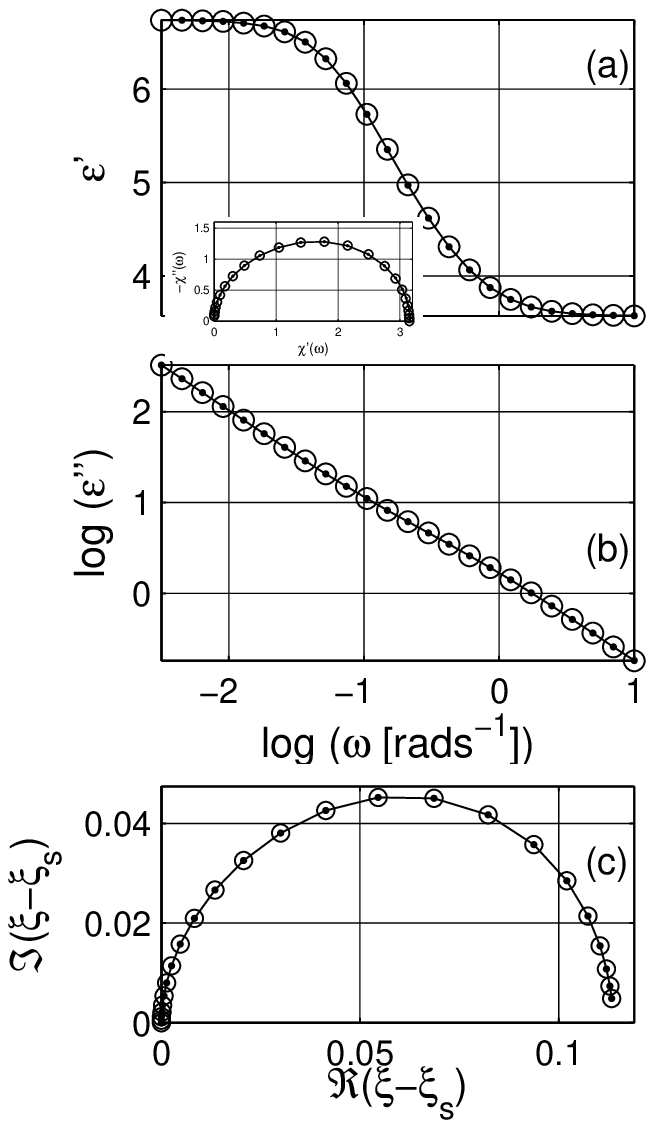}
  \caption{(a) The real and (b) the imaginary parts of the dielectric permittivity calculated with Eq.~(\ref{eq:LLL}) for $q=0.3$ and the modeled response obttained with the application of the Havriliak-Negami expression to the scaled permittivity. The permittivities of the matrix and the inclusion phases are the same as in Fig.~\ref{fig:dataLLL3}. (c) Argand diagram of scaled permittivity without out the percolation strength contribution.  The inset in (a) shows the dielectric susceptibility after the substruction of ohmic conductivity and the permittivity at high frequencies.  The lines with points are the simulated data and the symbols ($\circ$) are the data estimates of the Havriliak-Negami expression, Eq.~\eqref{eq:1}. There is a very good agreement between simulated and analyzed data sets.  Logarithmic scale is base 10.}
  \label{fig:dataLLL3_HN}
\end{figure}
\begin{figure}[t]
  \centering
  \includegraphics[width=\figsize\linewidth]{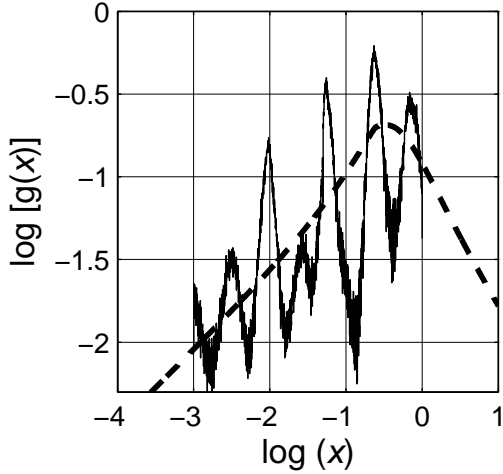}
  \caption{Comparison of the distributions ${sf g}$ obtained by the Havriliak-Negami (\kesik) and the numerical approach (\full). Observe that the Havriliak-Negami distribution also considers spectral parameter $x$ values larger than 1, $\log(x)>0$.   Logarithmic scale is base 10.}
  \label{fig:HNSDRdist}
\end{figure}
\begin{figure}[t]
  \centering
  \includegraphics[width=\figsize\linewidth]{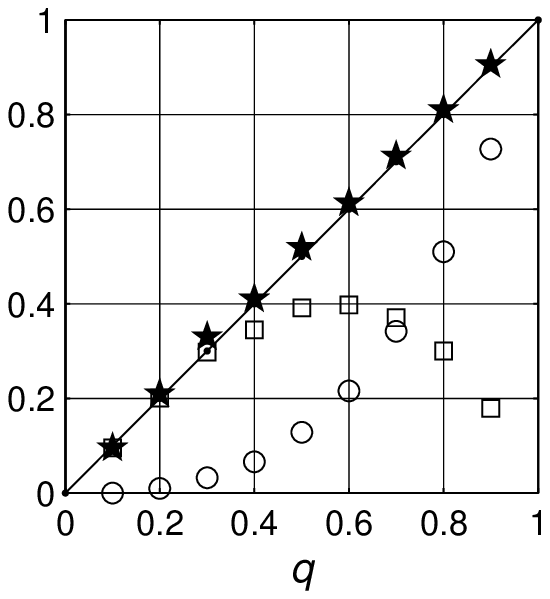}
  \caption{The most probable values of constants, $\overline{q}$ and $\overline{\xi_s}$ from the application of the Havriliak-Negami expression. The integral of ${\sf g}(x)$ yields the concentration of the inclusions $\overline{q}$, denoted with $(\Box)$ symbols. The percolation strength is $\overline{\xi}$, denoted with ($circ$) sysmbols. The sum of the two constants, $\overline{\xi_s}+\overline{q}$, denoted with ($\bigstar$), leads to the actual concentration, which is denoted with the solid line (\full).}
  \label{fig:LLL_Aandg_HN}
\end{figure}

Previously, it has been stated that the scaled permittivity $\xi$ in Eq.~(\ref{eq:xi}) can be expressed as in the convensional form as in the case of  dielectric relaxation \cite{Tuncer2005JPCMLET}, Eq.~(\ref{eq:1}). In order to show and verify this statement we apply a complex nonlinear least-squares curve fit algorithm to the scaled permittivity of the  LLL expression, denoted as $\xi^{\rm LLL}$ below. The error in the curve fitting procedure is used to quantify the fitness of the model function in Eq.~(\ref{eq:1}). The error is calculated as the sum of relative error at each point as follows,
\begin{eqnarray}
  \label{eq:3}
  \mathfrak{E}=\sum\left[
    \frac{\Re(\xi^{\rm LLL})-\Re(\xi^{\rm HN})}{\Re(\xi^{\rm LLL})}\right]^2+
\left[
    \frac{\Im(\xi^{\rm LLL})-\Im(\xi^{\rm HN})}{\Im(\xi^{\rm LLL})}\right]^2
\end{eqnarray}
Here $\xi^{\rm HN}$ is the model expression of Eq.~(\ref{eq:1}). The fit results are listed in Table~\ref{tableHN}, where the model values for the spectral parameter $x$, concentration $q$ and the percolation strength $\xi_s$ are presented with over-lines as in the previous section.

\begin{table}
\caption{Fit parameters of Havriliak-Negami\cite{HN} expression to scaled permittivity, Eq.~(\ref{eq:1}). The error $\mathfrak{E}$ is calculated with Eq.~(\ref{eq:3}).\label{tableHN}}
\begin{tabular}{crrrrr||r}
\hline\hline
$q$ & $\alpha$   &  $\beta$  & $\overline{x}$ &$\overline{q}$& $\overline{\xi_s}$ & \multicolumn{1}{c}{$\mathfrak{E}$}\\
\hline\hline
   0.1 &   1.034 &   0.438 &   0.536 &   0.095 &   0.000 &1.19$\times10^{-3}$ \\
   0.2 &   0.822 &   0.555 &   0.514 &   0.200 &   0.009 &7.28$\times10^{-6}$ \\
   0.3 &   0.716 &   0.635 &   0.464 &   0.297 &   0.032 &2.45$\times10^{-4}$ \\
   0.4 &   0.850 &   0.454 &   0.522 &   0.345 &   0.066 &4.27$\times10^{-6}$ \\
   0.5 &   0.800 &   0.470 &   0.494 &   0.391 &   0.128 &2.06$\times10^{-5}$ \\
   0.6 &   0.828 &   0.422 &   0.493 &   0.397 &   0.216 &4.08$\times10^{-6}$ \\
   0.7 &   0.822 &   0.409 &   0.475 &   0.371 &   0.342 &5.20$\times10^{-6}$ \\
   0.8 &   0.815 &   0.399 &   0.456 &   0.300 &   0.510 &6.41$\times10^{-6}$ \\
   0.9 &   0.810 &   0.390 &   0.438 &   0.179 &   0.727 &7.60$\times10^{-6}$ \\
\hline\hline
\end{tabular}
\end{table}

The fit results are shown in Fig.~\ref{fig:dataLLL3_HN} for $q=0.3$ case in the similar form as Fig.~\ref{fig:dataLLL3}. There is actually no particular differences between the two methods, except that the numerical techniques based on the Monte Carlo algorithm is capable of resolving individual peaks, as shown in the comparison graph in Fig.~\ref{fig:HNSDRdist}. It is not clear for example from the Havriliak-Negami approach that the system indicate a self-similar fractal-like structure. One should note that the Havriliak-Negami distribution, cf.~\S~\ref{sec:havr-negami-distr} \cite{Tuncer2000b}, presented below, even spread over spectral parameter values larger than one, $x>1$, which is not possible in the spectral density representation. However, in order to analyze the data it is very convenient and trivial to implement in avaliable curve fitting programs. Similar to Fig.~\ref{fig:LLL_Aandg},  the estimated concentration $\overline{q}$ and percolation strength $\overline{\xi_s}$ from the parametric analysis satisfy the condition in Eq.~\eqref{eq:A51} as shown in Fig.~\ref{fig:LLL_Aandg_HN}.




\section{Conclusions}\label{conclusions}
In this paper, we first derived an expression for the dielectric mixtures that resembles the distribution of relaxation times representation in dielectric relaxation phenomenon. In the derivation we used the spectral density representation. It is shown that the extisting knowledge on the dielectric relaxation theory can be applied to the dielectric properties of composites. In order to confirm the hypothesis, both a method similar to estimate the distribution of relaxation times and an extensively used empirical formula to express dielectric relaxation are employed to estimate the spectral density functions of composites simulated with the LLL expression. 
The numerical method based on the Monte Carlo technique estimated couple of peaks which did not change their location in the spectra with increased concentration of inclusions. This static behavior of the spectra resembles that there exists a hierarchical structural order in the composite, as a result, we infer that the LLL expression is proper for systems with self-similar fractal nature, such as composites with colloid aggregates and porous materials. We have explicitely shown why the LLL expression could be applied to describe the dielectric properties of powdered and porous systems. 
Last but not least, the findings are significant to confirm the structure of composite systems, whose dielectric permittivities are described with the LLL expression, in the literature. 


\appendix
\section{Derivation of the simple form}
\label{sec:appendix}
Eq.~(\ref{eq:generalwiener_e}) is expanded as follows,
\begin{eqnarray}
  \label{eq:A1}
  \frac{\vare-\varm}{\varm}=q\ A \left(\frac{\vare-\varm}{\varm}\right)+q\int_0^1\frac{{\sf G}(x)\dd x}{\left(\frac{\displaystyle\vare-\varm}{\displaystyle \varm}\right)^{-1}+x}
\end{eqnarray}
Let $\varepsilon_i-\varepsilon_j\equiv\Delta_{ij}$, then,
\begin{eqnarray}
  \label{eq:A2}
  \frac{\Delta_{\sf em}}{\varm}=q\ A \frac{\Delta_{\sf im}}{\varm}+\int_0^1\frac{q{\sf G}(x)\Delta_{\sf im} \dd x}{\varm+\Delta_{\sf im}x}
\end{eqnarray}
Now, multiply both sides with $\varm$,
\begin{eqnarray}
  \label{eq:A3}
  {\Delta_{\sf em}}=q\ A {\Delta_{\sf im}}+\int_0^1\frac{q{\sf G}(x)\Delta_{\sf im} \dd x}{1+\varm^{-1}\Delta_{\sf im}x}
\end{eqnarray}
Finally, let $\Delta_{\sf em}/\Delta_{\sf im}\equiv\xi$ and $qA\equiv\xi_s$, we obtain Eq.~(\ref{eq:spectral}). The properties of ${\sf G}$ and $A$ are such that\cite{Ghosh,GhoshFuchs,Gonc2000},
\begin{eqnarray}
  \label{eq:A4}
  A+\int_{0^+}^{1} {\sf G}(x) \dd x &=&1,\\
  \int_{0^+}^{1} x {\sf G}(x) \dd x &=&\frac{\displaystyle q(1-q)}{\displaystyle d}.
\end{eqnarray}
Here, $d$ is the dimension of the system. When we consider our new notation then,
\begin{eqnarray}
  \xi_s+\int_{0^+}^1 {\sf g}(x) \dd x &=&q,  \label{eq:A51}
\\
  \int_{0^+}^1 x {\sf g}(x) \dd x &=&\frac{\displaystyle (1-q)}{\displaystyle d}.  \label{eq:A52}
\end{eqnarray}

\section{Havriliak-Negami distribution function}
\label{sec:havr-negami-distr}
Havriliak and Negami~\cite{HN} have combined the works of Cole and Cole~\cite{CC} and Davidson and Cole~\cite{CD} and have expressed the dielectric dispersion with an asymmetric formula as presented in Eq.~(\ref{eq:1}). When $\alpha=\beta=1$, Eq.~(\ref{eq:1}) becomes the simple Debye and Maxwell Garnett equations for dielectrics or dielectric mixtures, respectively. Other interesting cases are when $\alpha=1$; Davidson-Cole expression and when $\beta=1$; Cole-Cole expression. Havriliak and Negami~\cite{HN2} have used the distribution of relaxation times as expressed by Davidson and Cole~\cite{CD}, and have substituted $s\exp(\pm\imath\pi)$ for $\imath s$. The solution for the distribution relaxation times and the spectral density function ${\sf g}(s)$, then, becomes,
\begin{equation}
  \label{eq:havneg_dist}
  {\sf g}(s)=\frac{1}{\pi} \left | \frac{10^{s\alpha\beta}\sin(\beta\Theta)}{[10^{2s\alpha}+2 \cdot 10^{s\alpha}\cos(\alpha\pi)+1]^{\beta/2}} \right |
\end{equation}
where,
\begin{displaymath}
 \label{eq:havnew_theta}
  \Theta=\arctan \left [ \frac{\displaystyle \sin(\alpha\pi)}{\displaystyle 10^{s\alpha}+\cos(\alpha\pi)} \nonumber \right ] 
\end{displaymath}
and $s=\log(x/\overline{x})$ with $\overline{x}$ being the most probable spectral parameter.


\end{document}